\documentclass[sigplan,screen]{acmart}

\setcopyright{acmlicensed}
\acmPrice{15.00}
\acmDOI{10.1145/3372885.3373836}
\acmYear{2020}
\copyrightyear{2020}
\acmISBN{978-1-4503-7097-4/20/01}
\acmConference[CPP '20]{Proceedings of the 9th ACM SIGPLAN International Conference on Certified Programs and Proofs}{January 20--21, 2020}{New Orleans, LA, USA}
\acmBooktitle{Proceedings of the 9th ACM SIGPLAN International Conference on Certified Programs and Proofs (CPP '20), January 20--21, 2020, New Orleans, LA, USA}

\usepackage{booktabs}
\usepackage{subcaption}
\usepackage{amsthm}
\usepackage{stmaryrd}
\usepackage{mathrsfs}
\usepackage{mdframed}
\usepackage{wrapfig}
\usepackage{bussproofs}
\usepackage{balance}
\newenvironment{bprooftree}
  {\leavevmode\hbox\bgroup}
  {\DisplayProof\egroup}

\usepackage{listings}
\lstdefinelanguage{PVS}{
   morekeywords={AND, CONJECTURE, FACT, LET, TABLE, ANDTHEN, CONTAINING, FALSE, LIBRARY, THEN, ARRAY, CONVERSION, FORALL, MACRO, THEOREM, ASSUMING, CONVERSION+, FORMULA, MEASURE, THEORY, ASSUMPTION, CONVERSION-, FROM, NONEMPTY, TYPE, TRUE, AUTO, REWRITE, COROLLARY, FUNCTION, NOT, TYPE, AUTO, REWRITE+, DATATYPE, HAS, TYPE, O, TYPE+, AUTO, REWRITE-, ELSE, IF, OBLIGATION, VAR, AXIOM, ELSIF, IFF, OF, WHEN, BEGIN, END, IMPLIES, OR, WHERE, BUT, ENDASSUMING, IMPORTING, ORELSE, WITH, BY, ENDCASES, IN, POSTULATE, XOR, CASES, ENDCOND, INDUCTIVE, PROPOSITION, CHALLENGE, ENDIF, JUDGEMENT, RECURSIVE, CLAIM, ENDTABLE, LAMBDA, SUBLEMMA, CLOSURE, EXISTS, LAW, SUBTYPES, COND, EXPORTING, LEMMA, SUBTYPE, OF},
    sensitive=false, 
    morecomment=[l]{\%}, 
    morestring=[b]" 
}
\definecolor{gainsboro}{rgb}{0.86, 0.86, 0.86}
\newcommand{\lstin}[1]{\lstinline[language=PVS]{#1}}

\begin{document}

\title{A Verified Packrat Parser Interpreter for Parsing Expression
  Grammars}
\titlenote{This work was supported by the National Institute of Aerospace
 Award C18-201097-SRI, NSF Grant
 SHF-1817204, Ecole Polytechnique, and DARPA under agreement number HR001119C0075. 
 The views and conclusions contained herein are those of the authors and
 should not be interpreted as necessarily representing the official
 policies or endorsements, either expressed or implied, of NASA, NSF, DARPA, Ecole Polytechnique, or the
 U.S. Government.  We thank the anonymous referees for their detailed comments and 
 constructive feedback.
}

\author{Clement Blaudeau}
 \affiliation{
 \institution{Ecole Polytechnique}
 \streetaddress{Route de Saclay}
 \city{Palaiseau Cedex}
 \postcode{91128}
 \country{France}}
\email{clement.blaudeau@polytechnique.edu}

\author{Natarajan Shankar}
 \affiliation{
 \department{Computer Science Laboratory}
 \institution{SRI International}
 \streetaddress{333 Ravenswood Avenue}
 \city{Menlo Park}
 \state{CA}
 \postcode{94025}
 \country{USA}}
\email{shankar@csl.sri.com}

\begin{abstract}
  Parsing expression grammars (PEGs) offer a natural opportunity for
  building verified parser interpreters based on higher-order parsing
  combinators.  PEGs are expressive, unambiguous, and efficient to
  parse in a top-down recursive descent style.  We use the rich type
  system of the PVS specification language and verification system to
  formalize the metatheory of PEGs and define a reference
  implementation of a recursive parser interpreter for PEGs.  In order
  to ensure termination of parsing, we define a notion of a
  well-formed grammar.  Rather than relying on an inductive definition
  of parsing, we use abstract syntax trees that represent the
  computational trace of the parser to provide an effective proof
  certificate for correct parsing and ensure that parsing properties
  including soundness and completeness are maintained.  The
  correctness properties are embedded in the types of the operations
  so that the proofs can be easily constructed from local proof
  obligations.    Building on
  the reference parser interpreter, we define a packrat parser interpreter
  as well as an extension that is capable of semantic interpretation.  Both
  these parser interpreters are proved equivalent to the reference one.
  All of the parsers are executable.  The proofs are formalized in
  mathematical terms so that similar parser interpreters can be defined
  in any specification language with a type system similar to
  PVS. \end{abstract}



\maketitle


\section{Introduction}\label{sec:intro}

\paragraph{}
Parsing is the process of extracting structure and information from a
string of tokens according to a formal
grammar~\cite{dragonbook}\@. For critical applications,
parsing errors and lack of proper input validation can be a common
source of vulnerability and inconsistency leading to numerous errors
and attacks~\cite{DBLP:journals/usenix-login/BratusHHLMPS17}\@. 
Parser issues can come from the failure to follow a formal grammar,
and from the formal grammar itself. Grammars might, by design or
accident, introduce complexity, nontermination, and nondeterminism.
For example, the \emph{dangling else} problem where both
\textsf{if A then B} and \textsf{if A then B else C} are well-formed, 
leads to an ambiguity in parsing an expression of the form
\textsf{if A then if B then C else D} where the \textsf{else} branch
could be associated with either of the two conditionals.  
Ford~\cite{DBLP:conf/popl/Ford04} introduced Parsing
Expression Grammars (PEGs) as a formalism for defining unambiguous
grammars that can be parsed efficiently using a recursive descent
scheme with parsing combinators for the basic grammar operators~\cite{Burge,DBLP:conf/fp/Hutton89,DBLP:journals/cj/FrostL89}\@.  We present a PVS~\cite{Owre95:prolegomena} formalization
of PEGs with a rigorous treatment of well-formed grammars and a
reference parser generator for such grammars.  We also verify a
packrat parser generator for PEGs and demonstrate that it is equivalent
to the reference parser generator.  Our parse tree
representation can serve as an independent proof-of-parse for PEGs making it possible to
validate the result produced by the correct parsing, successful or unsuccessful,
of an input string with reference to a given grammar regardless of how
the parsing was performed.\footnote{The  PVS formalizations can be accessed at \url{https://github.com/SRI-CSL/PVSPackrat}\@. }

Parsing expression grammars were introduced as a pragmatic
compromise between efficiency and expressiveness.  These grammars are
similar to context-free grammars in supporting terminal symbols,
concatenation $A ; B$, and iteration $A^*$, but the
choice operation  $A \mid B$ is
replaced with a priority operation $A / B$ where the grammar $B$ is
matched against the input only when the match on $A$ fails.  PEGs also include
option ($?$), test ($\&$), and negation operations ($!$):
$?A$ either matches (and consumes) some prefix of
the current input or succeeds without consuming any tokens; 
$\&A$ tests if there is a prefix of the input
that matches $A$ without consuming any tokens, and $!A$ fails if 
some input prefix matches $A$\@.  PEGs are unambiguous:
there is at most one way to parse the input with respect to a given
grammar.  PEGs are greedy in that compared to the context-free analog,
a PEG grammar represents the longest parse.  For example, parsing an
input
with respect to $A^*$ will consume as many instances of $A$ from the
input as possible\@.   The test operations can be used to define test
predicates that can look ahead into arbitrarily long prefixes of the
input without actually consuming any input, 
making it possible to even capture certain
non-context-free grammars such as $a^nb^nc^n$\@.

A PEG grammar consists of a set of productions mapping  nonterminals
to PEG expressions in the nonterminals.  Not all PEGs are
well-formed.  For example, $A^*$ is not well-formed when $A$ 
can match the empty string.  A left-recursive grammar like
$A \rightarrow A/a$ is not well-formed since the corresponding
unfolding of the grammar might not converge.  

The parsing of well-formed PEGs can be directly implemented by a
recursive descent parser in which each construct is supported by a
parsing combinator.  For example, the choice construct $A / B$ is
implemented by a combinator that takes a parser for $A$ and a parser
for $B$, and invokes the parser for $B$ on the input only when the
parser for $A$ returns with failure.  This kind of parsing scheme can
have exponential complexity through repeated parsing queries with the
same nonterminal.  Consider the following grammar~\cite{DBLP:conf/icfp/Ford02}:
\begin{eqnarray*}
  A &::=& M + A / M\\
M &::=&  G * M / G\\
G &::=& (A) / \mathit{int}
\end{eqnarray*}

Parsing an input, say $(((5 * 4) * 3) * 2) * 1$ requires $2^4$
parser calls for the nonterminal $M$.  This duplication is avoided in
packrat parsing by memoizing the results of the parse.

PEGs and PEG parsing is defined in Section~\ref{sec:peg}
where we also discuss the termination problem for parsing with respect to
arbitrary PEGs.  In Section~\ref{sec:wfpeg}, we outline the well-formedness
properties of PEGs that guarantee the termination of parsing.
We present the verification of three parser
interpreters for PEGs.    
The first is a reference parser interpreter for PEG
grammars presented in Section~\ref{sec:refparser}.  The second
is a packrat
parser interpreter described in Section~\ref{sec:packrat}.
Section~\ref{sec:semantic} describes 
a variant  that allows the output parse tree to
be customized according to semantic actions.  Section~\ref{sec:conclusions}
makes some brief concluding observations.  

\paragraph{Related Work.}
Though parsing is an important firewall between untrusted data and
vulnerable applications, there have only been a few instances of
verified parsers or parser interpreters/generators.  The parser front-end was one
of the few identifiable weaknesses of the verified CompCert
compiler~\cite{journals/cacm/Leroy09}\@.  
Barthwal and Norrish~\cite{conf/esop/BarthwalN09} present 
an SLR parser generator verified in HOL4 that produces an independently verifiable
parse tree.  Ridge~\cite{conf/cpp/Ridge11} has verified
the termination, soundness, and completeness of a recursive descent parser based on
parsing combinators for context-free languages.  The RockSalt checker
of Morrisett, Tan, Tassarotti, Tristan, and Gan~\cite{DBLP:conf/pldi/MorrisettTTTG12} for
checking software-based fault isolation of native executable
code in a browser employs a regular expression parser for x86 instructions
that has been verified to be sound in Coq. In subsequent work, Tan
and Morrisett~\cite{DBLP:journals/jar/TanM18} certified
encoder/decoder pairs are constructed from bidirectional grammars. 
Lopes, Ribeiro, and Camar\~ao~\cite{DBLP:conf/sblp/LopesRC16} have
also verified a regular expression parser using Idris~\cite{DBLP:journals/jfp/Brady13}\@.  
Bernardy and Jansson~\cite{DBLP:journals/corr/BernardyJ16}
have formalized Valiant's algorithm for parsing context-free languages in Agda~\cite{DBLP:conf/tphol/BoveDN09}\@. 
Koprowski and Binsztok~\cite{trx} present a Coq
verification of a parser interpreter called TRX for PEGs.
The CakeML compiler which has been verified using HOL4 employs
a similar verified PEG parser interpreter~\cite{DBLP:conf/popl/KumarMNO14}\@.  
Lasser, Casinghino, Fisher, and
Roux~\cite{DBLP:conf/itp/LasserCFR19} have verified an LL(1) parser
generator covering the generation of the lookup table and the
stack-based parser.  
Jourdan, Pottier, and
Leroy~\cite{conf/esop/JourdanPL12} define a generator for an independently
and verifiably validatable parsing automaton for an LR(1) grammar.
A number of papers address the verification and synthesis of  decoder/encoder
pairs~\cite{DBLP:conf/icfp/GeestS17,DBLP:conf/cav/TullsenPCT18,ramananandro2019everparse,DBLP:journals/pacmpl/DelawareSPYC19,DBLP:conf/cpp/YeD19} where the objective is to serialize and deserialize data. 

Of the related projects, the TRX parser~\cite{trx}
and the CakeML parser ~\cite{DBLP:conf/popl/KumarMNO14} are 
closest to the one presented here.  We build on the TRX work,
particularly in the treatment of termination for PEGs and in the
definition of the non-packrat PEG parser.  However, we define an
executable check for grammar well-formedness in contrast to the
inductive characterization in TRX.  We also present a parse
tree representation of both successful and failed parses.
We show that these parses are unique for a given grammar thus
capturing both the soundness and completeness of the parser interpreter.  The TRX
verification only captures the soundness argument using an inductively
defined parse derivation that captures the parse semantics.  While
parse trees have been used as proofs-of-parse, in the
work of Barthwal and Norrish~\cite{conf/esop/BarthwalN09} and Ridge~\cite{conf/cpp/Ridge11},
we have extended it to capture both success and failure.  We also
go beyond the reference PEG parser interpreter to verify two packrat
parser interpreters, one without semantic actions and one with.  The PVS
proofs we present take advantage of predicate subtyping in PVS to
craft a verification methodology that reduces the correctness of the
grammar analyzer and the parser interpreters to small, local proof
obligations with easy automated proofs instead of big theorems with
manually generated proof structures.  Our proof methodology is
transferable to other parsing formats and parser generation algorithms.

\section{Parsing Expression Grammars}\label{sec:peg}

Parsing Expression Grammars (PEGs) were introduced by Ford as a
formalism for capturing grammars that correspond to greedy, unambiguous recursive
descent parsing.  To describe PEGs, we rely on operators 
resembling those in context-free grammars, with the notable
exception of the prioritized choice operator. Here is a formal definition of
PEGs, relying on two types: \(V_T\), the type of terminals (in most
cases bytes or characters, but applicable to any type instantiating \(V_T\)) and \(V_N\), the type of nonterminals that
are basically patterns: the set of grammar expressions \(\Delta\) is
inductively defined as below.  
\begin{equation}
	\begin{array}{rccll}
	\Delta&::= &\epsilon    &\text{empty expression}&              \\
	&|	& [\cdot]     &\text{any character}   &              \\
	&|	& [a]         &\text{a terminal}      & (a\in V_T)   \\
	&|	& A           &\text{a nonterminal}  & (A \in V_N)  \\
	&|	& e1;e2       &\text{a sequence}      & (e1, e2 \in \Delta)  \\
	&|	& e1/e2       &\text{a prioritized choice}  & (e1, e2 \in \Delta)  \\
	&|	& e*          &\text{a greedy repetition}   & (e \in \Delta)  \\
	&|	& !e          &\text{a not-predicate}       & (e \in \Delta)  \\
	\end{array}
\end{equation}
		In addition to those basic operators, we have a few
                other operators that can be emulated by the basic ones:
		\begin{enumerate}
			\item $[a-z]$ the range, equivalent to $[a] / [b] / ... / [z]$
			\item $[``s"]$ the string, equivalent to
                          $[c_1];[c_2];...;[c_n]$, where $s =
                          c_1c_2\ldots c_n$ for tokens $c_1,\ldots, c_n$
			\item $e+$ the plus operator, equivalent to $e;e*$
			\item $e?$ the optional operator, equivalent to $e/\epsilon$
			\item $\&e$ the and operator, equivalent to $!!e$
		\end{enumerate}
		In practice these additional operators would certainly
                be used, but as they can be emulated by the
                basic ones, we can ignore them for the proofs in order
                to avoid redundant cases during the case analyses.
	
			\begin{figure}[h!]
			\label{peg_pvs}
\begin{lstlisting}[language = PVS]
peg: DATATYPE
BEGIN
	ε:  ε?
	any: any?
	terminal(a: V_T): terminal?
	nonTerminal(A: V_N): nonTerminal?
	seq(e1, e2: peg): seq?
	prior(e1, e2: peg): prior?
	star(e: peg): star?
	notP(e: peg): notP?
END peg
\end{lstlisting}
			\caption{PVS code for the PEG grammars formalized as an algebraic datatype with constructors, accessors, and recognizers}
			\end{figure}
			
		\paragraph{Restriction to a Finite Set of
                  Nonterminals.} Technically we could consider $V_N$
                as any set, finite or not, and build results making no
                other assumptions. In practice however, $V_N$ is the
                set of patterns given by the creator of the
                grammar. So we can consider $V_N$ finite and bounded
                by $n + 1 = Card(V_N)$. In the following,   $V_N$
                and $\llbracket 0, n \rrbracket$ are used interchangeably.
                
		\paragraph{The Problem of Termination.}
			Though PEGs are 
                        unambiguous~\cite{DBLP:conf/popl/Ford04}, 
                        this does not ensure that parsing
                        terminates. The most basic non-terminating
                        PEG expression is $\epsilon *$ since parsing with it loops
                        forever without consuming any input.  It shows that greedy operators
                        should not rely on expressions that can
                        \emph{succeed without consuming tokens from
                          the input}. Also, the
                        use of nonterminals can easily introduce non-terminating left-recursion.
                        For example, with $V_N =
                        \{A,B\}$ and grammar production map $P_{exp}$
                        from $V_N$ to $\Delta$ defined as:
			\[
				\begin{array}{rl}
					P_{exp}(A) &= B \\
					P_{exp}(B) &= A				
				\end{array}
			\]
			The parsing here would loop forever.
                        This introduces the two main considerations regarding the termination of parsing: consumption of characters and prevention of infinite left-recursion.  In the following section, we
                        define the properties of a \emph{well-formed} grammar that ensure termination of parsing regardless of the input. 
\section{Well-Formed PEG Grammars}\label{sec:wfpeg}
We develop a new approach to well-formed PEGs with a
computational point of view.  Our focus is on constructive definitions
that yield easy implementations. The notions of structural
well-formedness and pattern well-formedness are, along with the
use of abstract proof-of-parse trees, the major
differences with respect to the verified PEG parser TRX~\cite{trx}\@.  
		\subsection{Consumption and Grammar Properties}
			To characterize terminating grammars, we first need to identify the relevant properties of grammar expressions that ensure termination.  We then need to compute them. The most basic property is  \emph{success without consumption} (of tokens). However, the not-predicate, which succeeds if the expression fails, requires us to also know if expressions can \emph{fail}. The last case is \emph{success with consumption} of at least one token of the input.  We introduce below a precise mathematical formalization of the notion of a grammar property that corresponds to the way it is designed and proved in PVS. 
			\paragraph{Formalization.}
			 To translate those properties, we define
                         three predicates: \[P_\bot, P_0,P_{>0}:
                           \Delta \longrightarrow \textit{bool},\]
representing whether a parse based on the grammar can fail,  succeed without
consuming input, or succeed only by consuming input, respectively.  We define
$P_{\geq 0}(e)$ as $P_0(e) \vee P_{>0}(e)$\@.  
			The inductive definition relies on the rules
                        shown in Figure \ref{prop_axioms}.
	\begin{figure*}
	\begin{center}
		\begin{bprooftree}
			\AxiomC{\vphantom{V}}
			\UnaryInfC{$P_0(\epsilon)$}
		\end{bprooftree}\enspace
		\begin{bprooftree}
			\AxiomC{\vphantom{V}}
			\UnaryInfC{$P_{>0}([\cdot])$}
		\end{bprooftree}\enspace
		\begin{bprooftree}
			\AxiomC{\vphantom{V}}
			\UnaryInfC{$P_{\bot}([\cdot])$}
		\end{bprooftree}\enspace
		\begin{bprooftree}
			\AxiomC{$a\in V_T$}
			\UnaryInfC{$P_{>0}([a])$}
		\end{bprooftree}\enspace
		\begin{bprooftree}
			\AxiomC{$a\in V_T$}
			\UnaryInfC{$P_{\bot}([a])$}
		\end{bprooftree}\enspace
		\begin{bprooftree}
			\AxiomC{$P_\bot(e)$}
			\UnaryInfC{$P_{0}(e*)$}
		\end{bprooftree}\enspace
		\begin{bprooftree}
			\AxiomC{$P_{>0}(e)$}
			\UnaryInfC{$P_{>0}(e*)$}
		\end{bprooftree}
	\end{center}\vspace{2mm}
	\begin{center}
		\begin{bprooftree}
			\AxiomC{$\star \in \{0, >0, \bot\}$}
			\AxiomC{$A\in V_N$}
			\AxiomC{$P_\star(P_{exp}(A))$}
			\TrinaryInfC{$P_{\star}(A)$}
		\end{bprooftree}\enspace
		\begin{bprooftree}
			\AxiomC{$P_{\bot}(e_1)\vee \left[ P_{\geq 0}(e_1)\wedge P_\bot(e_2) \right] $}
			\UnaryInfC{$P_{\bot}(e_1;e_2)$}
		\end{bprooftree}\enspace
	\end{center}\vspace{2mm}
	\begin{center}
		\begin{bprooftree}
			\AxiomC{$P_{0}(e_1)$}
			\AxiomC{$P_0(e_2)$}
			\BinaryInfC{$P_{0}(e_1;e_2)$}
		\end{bprooftree}\enspace
		\begin{bprooftree}
			\AxiomC{$\left[ P_{>0}(e_1) \wedge P_{\geq 0}(e_2)\right] \vee \left[ P_0(e_1) \wedge P_{>0}(e_2)\right]$}
			\UnaryInfC{$P_{>0}(e_1;e_2)$}
		\end{bprooftree}\enspace
	\end{center}\vspace{2mm}
	\begin{center}
		\begin{bprooftree}
			\AxiomC{$P_{0}(e_1) \vee \left[ P_\bot(e_1) \wedge P_0(e_2) \right]$}
			\UnaryInfC{$P_{0}(e_1/e_2)$}
		\end{bprooftree}\enspace
		\begin{bprooftree}
			\AxiomC{$P_{>0}(e_1) \vee \left[ P_{\bot}(e_1) \wedge P_{>0}(e_2) \right]$}
			\UnaryInfC{$P_{>0}(e_1/e_2)$}
		\end{bprooftree}\enspace
	\end{center}\vspace{2mm}
	\begin{center}
		\begin{bprooftree}
			\AxiomC{$P_\bot(e_1)$}
			\AxiomC{$P_\bot(e_2)$}
			\BinaryInfC{$P_\bot(e_1/e_2)$}
		\end{bprooftree}\enspace
		\begin{bprooftree}
			\AxiomC{$P_\bot(e)$}
			\UnaryInfC{$P_0(!e)$}
		\end{bprooftree}\enspace
		\begin{bprooftree}
			\AxiomC{$P_{\geq 0}(e)$}
			\UnaryInfC{$P_\bot(!e)$}
		\end{bprooftree}
	\end{center}
	\caption{Inductive rules for grammar properties}
	\label{prop_axioms}
	\end{figure*}

			This inductive approach is valid and matches the intuition, and can be easily formalized in PVS. However, we want an effective computational mechanism for checking the tree structure of a grammar expression that might contain loops\footnote{With $A$ relying on the properties of $B$ and vice-versa}. We can tackle this problem as follows: we first introduce the set of \emph{already known properties of nonterminals}, that starts as the empty set. Then we try to compute all the new properties that can be obtained with current knowledge, and iterate until no new properties are found. The number of computed properties of nonterminals is non-decreasing, and bounded by $3\times |V_N|$. When all the properties  of each nonterminal are known, we can compute in a straightforward recursive way the properties of any grammar involving these nonterminals.  We formalize this approach below by first introducing a function over the nonterminals: 
			$P:V_N\longrightarrow \{\text{known},\; \text{unknown}\}^3$ that represent the known properties. Let $\mathbb{P}$ be the set of those predicates. In the actual implementation, we use \lstin{bool} for representation of \emph{known/unknown}. It is important to notice that in our model, a \emph{false} does not mean that the property \emph{is} false, but that it is \emph{unknown}, so that only \emph{true} yields useful information. 
			We introduce an order:
			 \[
			 \forall P,P' \in \mathbb{P}, \; P\leq P' \Rightarrow \forall A \in V_N
			 \left\{ 
			 \begin{array}{l} 
				P(A)(1) \Rightarrow P'(A)(1) \\
				P(A)(2) \Rightarrow P'(A)(2) \\
				P(A)(3) \Rightarrow P'(A)(3) 
			\end{array} \right. 
			\]
			This order is trivially reflexive and transitive. We can now introduce a function that computes the properties of a grammar node based on current knowledge: the code is given in Figure~\ref{grammar_props}.
			\[
				g: \Delta \times \mathbb{P} \longrightarrow [bool]^3			
                              \]
where \[
	\begin{array}{rl}
		P_\bot: \Delta \longrightarrow bool &= g(\cdot, P)(1) \\
		P_0: \Delta \longrightarrow bool &= g(\cdot, P)(2) \\
		P_{>0}: \Delta \longrightarrow bool &= g(\cdot, P)(3)	
	\end{array} \]
                              
			\begin{figure}[!ht]
\begin{lstlisting}[language=PVS]
% Recursively computes the grammar properties of a peg object
% based on a current set of known properties for non terminals
g_props(G, P) : RECURSIVE [bool, bool, bool] =
  CASES G of
    ε: (false, true, false),
    any: (true, false, true),
    terminal(a): (true, false, true),
    nonTerminal(A) : (P`1(A), P`2(A), P`3(A)),
    seq(e1, e2): LET (e1_f, e1_0, e1_s) = g_props(e1, P) IN
        LET (e2_f, e2_0, e2_s) = g_props(e2, P) IN
        (e1_f ∨ ((e1_0 ∨ e1_s) ∧ e2_f),
        e1_0 ∧ e2_0,
        (e1_s ∧ (e2_0 ∨ e2_s)) ∨ (e1_0 ∧ e2_s)),
    prior(e1, e2): LET (e1_f, e1_0, e1_s) = g_props(e1, P) IN
        LET (e2_f, e2_0, e2_s) = g_props(e2, P) IN
        (e1_f ∧ e2_f, 
        e1_0 ∨ (e1_f ∧ e2_0), 
        e1_s ∨ (e1_f ∧ e2_s)),
    star(e): LET (e_f, e_0, e_s) = g_props(e, P) IN
        (false, e_f, e_s),
    notP(e): LET (e_f, e_0, e_s) = g_props(e, P) IN
        (e_s ∨ e_0, e_f, false),
  ENDCASES
MEASURE pegMeasure(G)
\end{lstlisting}
			\caption{PVS implementation of the $g$ function defined by case analysis. PVS allows the use of the unicode characters for \emph{and, or, implies, iff}.}
			\label{grammar_props}
			\end{figure}

			This structurally recursive function satisfies  the rules from
                        Figure~\ref{prop_axioms}\@.  The termination
                        measure \texttt{pegMeasure(G)} is just the
                        size of \texttt{G} given by the number of
                        nodes.  
                        We also introduce a function $\rho$ that takes a nonterminal $A$ and a set of properties $P$, compute the properties of $A$ and returns $P$ extended with the new computed properties:
			\[
			\begin{array}{rll}
				\rho : V_N \times \mathbb{P} & \longrightarrow & \mathbb{P} \\
				               (A,P) & \longmapsto & \left\{ \begin{array}{l}
				               	   \rho(A,P)(A) = g(P_{exp}(A), P)\\
				                   \rho(A,P)(B) = P(B) \; \;(B\neq A)
				               \end{array} \right.			
			\end{array}
			\]
			Basically, we want to apply this function
                        $\rho$ a certain number of times (at most $3
                        \times |V_N|$) to get all reachable
                        properties. But the problem is that this
                        function is not monotonic, for example, when
                        $P(A) = (\mathit{known}, \mathit{known}, \mathit{known})$ and $P_{exp}(A) =
                        \epsilon$, $\rho(A, P) < P$\@.  Thus, we need to restrain the set of properties to what we call \emph{coherent properties}: sets of properties that are not contradicted by themselves. We define that set as:
			\begin{equation}
			\mathbb{C} = \{P \in \mathbb{P} \; |\; \forall A \in V_N, \; P \leq \rho(A,P) \}
			\end{equation}
			Under this assumption of coherence, $\rho$ is monotonic.
			\begin{lemma}[$\rho$ is monotonic]
				\begin{equation}
					\label{grp_increasing}
					\forall P,P', G \in \mathbb{C}^2 \times \Delta , \; P \leq P' \Rightarrow \rho(G,P) \leq \rho(G, P')
				\end{equation}
			\end{lemma}
			
			In the PVS implementation, we define the $\rho$ function, as well as a \emph{coherent properties} type, and prove the monotonicity result.

	\subsection{Properties Computation}
		Now that we have a clear formalization of the computation of grammar properties, we can focus on the way all the properties are recursively computed. Again, the main idea is that we only need the properties of the nonterminals to  compute the properties of any grammar node with a simple recursive function.
		\paragraph{Example.}
		Consider the example with the following $P_{exp}$ function:
		\[
		\begin{array}{rl}
		P_{exp}(A) &= \;[a] \\
		P_{exp}(B) &= \;!A /C \\
		P_{exp}(C) &= \;!B ;A
		\end{array} \]
		\begin{enumerate}
			\item We have $P_\bot(A)$ and $P_{>0}(A)$ immediately.
			\item We then get $P_\bot(!A)$ and $P_0(!A) \Rightarrow P_0(B)$.
			\item This gives us $P_\bot(!B) \Rightarrow P_\bot(C)$.
			\item Combining $P_\bot(!A)$ and $P_\bot(C)$ we get $P_\bot(B)$.
		\end{enumerate} 
		In this example we can see that for computing the
                properties of $B$ we need properties of $C$ and
                vice-versa: the computation cannot be done in a single
                 pass. 
		\paragraph{Computation Process.}
		The computation process of the nonterminal properties is the following:\footnote{It is interesting to notice that far more optimal ways could be invented to tackle this computational problem, especially using graphs and memoization to prevent computing over and over again the same things. But first of all, the aim here is to have a verified solution, and secondly, this computation is only done once and for all, it does not affect the parsing. Thus the optimality is not the main focus here.}
		\begin{enumerate}
			\item Starting with the empty set of properties $0_\mathbb{C}$, we compute the properties of all the \emph{nonterminals}, one by one, augmenting the set of properties as new ones are found.
			\item Once that is done, we check if new properties have been found since the start of the nonterminal computation. If so, we restart the computation, and otherwise, we return the result.
		\end{enumerate}
		\paragraph{Formalization.}
		Next, we formalize the computation process.  We define a sequence
                that translates the computation of properties for all
                the nonterminals, one by one. The property set on which the computation is made is the superscript, and the nonterminal on which we are trying to compute new properties is written in subscript. So to recompute the new properties for all nonterminals between $0$ and $n$ we have: 
		\begin{equation}
		\left\{ \begin{array}{rl}
			r_A^P &= r_{A+1}^{\rho(A,P)}	 \\		
			r_{n}^P &= \rho(n,P)
		\end{array} \right.
		\end{equation}
		Lemma~\ref{lem:recompute} captures three useful properties entailed
                by this sequence. 
		
		\begin{lemma}[Recomputing nonTerminal properties increases knowledge]\label{lem:recompute}
		\begin{equation}
		\label{recompute_nt}
		\begin{array}{rrl}
			\forall P, A \in \mathbb{C}\times \llbracket 0, n \rrbracket,\;& P&\leq r_A^P \\
			\forall P, P', A \in \mathbb{C}^2 \times \llbracket 0, n \rrbracket, \;& P \leq P' \Rightarrow r_A^P &\leq r_A^{P'} \\
			\forall P, A \in \mathbb{C}\times \llbracket 0, n-1 \rrbracket, \;& r_{A+1}^P &\leq r_A^P
		\end{array}
		\end{equation}				
              \end{lemma}
              
		The  expected result of the computation is a set of properties that cannot be extended (because it already has all the reachable properties). We can define the set of properties that are such \emph{fixpoints} of {computation:}
		\begin{equation}
			\mathbb{F} = \left\{P \in \mathbb{C} \; | \; P = r_0^P\right\}
		\end{equation}
		Such a fixpoint can be reached in a bounded (by $3*(n+1)$) number of steps by repeatedly applying the function $\phi(P) = r_0^P$ starting with $0_\mathbb{C}$. We have the following {results:}
		\begin{theorem}[Fixpoint properties]
			Recomputing the properties of a nonterminal with a set of properties that is a fixpoint gives back the same result:
			\begin{equation}
				\forall (P,A) \in \mathbb{F}\times V_N, \; P = \rho(A,P)
			\end{equation}			 
                      \end{theorem}
			
			In the PVS implementation, we define the $r$ function, as well as a \emph{fixpoint properties} type. We prove the lemmas and the theorem. This gives us an effective way to compute the properties of a given set of nonterminals.

	\subsection{Well-Formed Grammars}
\begin{figure}[h!]
	\includegraphics[width=0.35\textwidth]{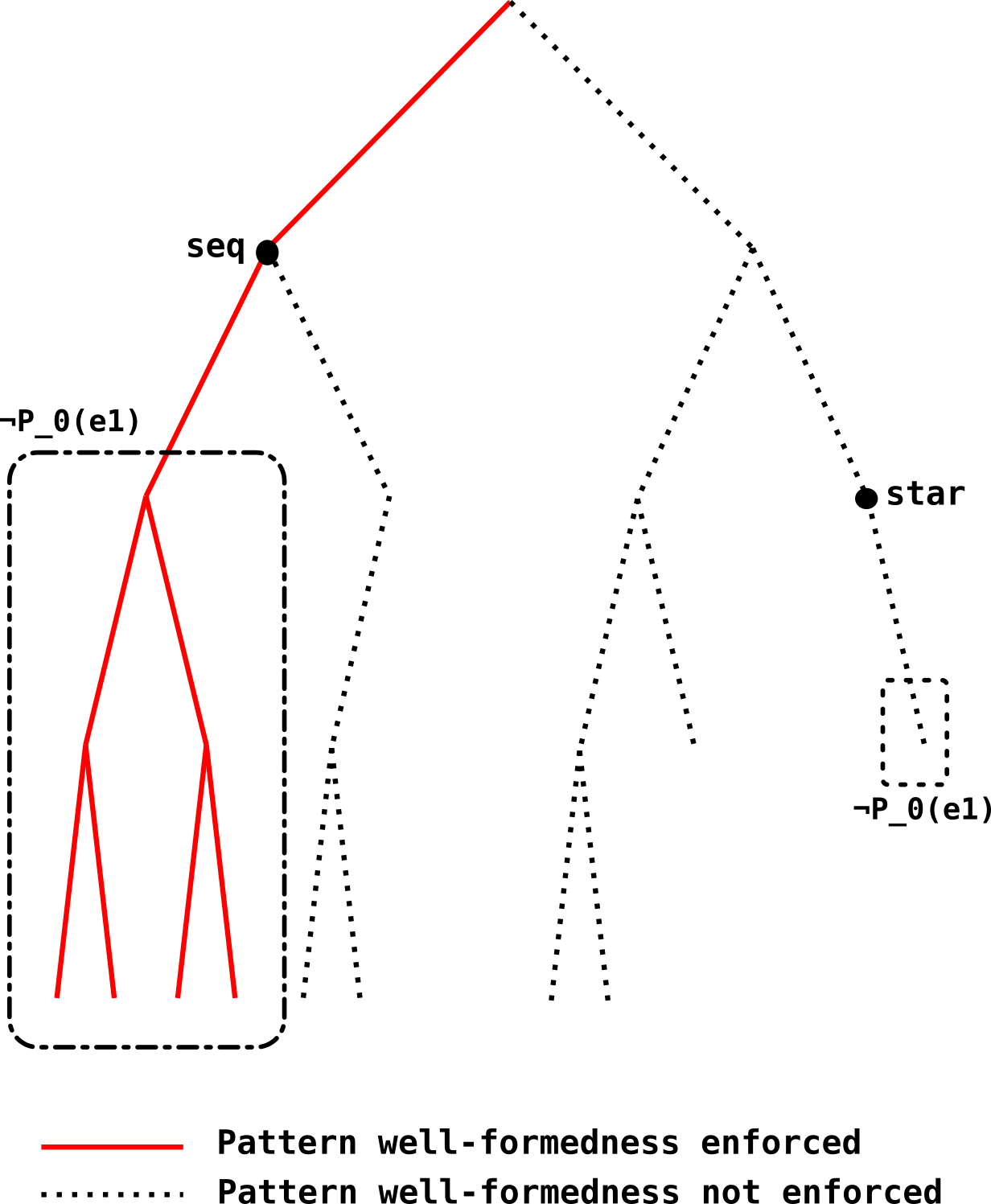}
	\caption{Representation of a well-formed grammar}
	\label{png:wfgrammar}
	\end{figure}

	Those properties allow us to define the grammar that we call \emph{well-formed}: grammars that structurally enforce  the termination of the parsing. The main idea is to prevent the two kinds of loops that we mentioned: structural ones ($\epsilon * $) and pattern ones ($P_{exp}(A) = B$, $P_{exp}(B) = A$). Two approaches are needed: 
	\begin{enumerate}
		\item Preventing \emph{structural aberrations} is
                  quite easy: we can just go through the whole tree
                  and check that every time the \emph{star} operator
                  is used, it is applied to a grammar node that cannot succeed without consuming any input. Thanks to the work previously done on properties computation, this is an easy task.
		\item Preventing \emph{pattern aberrations} is a bit
                  trickier, as we want to allow patterns to use other
                  patterns in some instances while  preventing them from doing so in others. 
                  The idea is the following: we assume
                  there exists an order\footnote{This approach is
                    actually equivalent to the inductive definition of
                    well-formedness that can be found in \cite{DBLP:conf/icfp/Ford02}:
                    if we have an order on nonterminals then the
                    inductive definition can follow that order, and
                    vice-versa}\footnote{For the scope of this paper, we assume that the user is able to provide the order for the nonterminals. We conjecture that if such an order actually exists, there exists ways to compute it that are more efficient than testing all possible orders. An approach using a graph of dependency between the nonterminals might be fruitful.} over the nonterminals where the
                  grammar $P_{exp}(A)$ for $A$ can only employ
                  strictly smaller nonterminals until it is clear that
                  at least one character is consumed.   For example, once the left-branch of a \emph{seq} is not of type $P_0$, the right branch can use any nonterminal.
	\end{enumerate}
	A visual representation is given in Figure \ref{png:wfgrammar}.
	
	\paragraph{Formalization.}
	Given those remarks we see that we can define a function $\omega$ that verifies both structural and pattern well-formedness of a grammar node that is a subterm of $P_{exp}(A)$ for a certain $A \in V_N$. The function always checks structural well-formedness, but pattern well-formedness is only checked on certain branches, using a special argument $\delta$ that is true if we enforce pattern well-formedness and that is false if it is not needed.

	\begin{equation}
		\begin{array}{rrll}
			\omega:& \Delta \times V_N \times bool & \longrightarrow & bool \\
			       & (\epsilon, A, \delta) & \longmapsto & \top \\
			       & ([\cdot], A, \delta) & \longmapsto & \top \\
			       & ([a], A, \delta) & \longmapsto & \top \\
			       & (B, A, \delta) & \longmapsto & \delta \Rightarrow (B <_{V_N} A) \\
			       & (e_1;e_2, A, \delta) & \longmapsto & \omega(e_1, A, \delta) \\
			       & & & \; \wedge \; \omega(e_2, A, \delta \wedge P_0(e_1)) \\
			       & (e_1/e_2, A, \delta) & \longmapsto & \omega(e_1,A,\delta) \wedge \omega(e_2, A, \delta) \\
			       & (e*, A, \delta) & \longmapsto & \omega(e, A, \delta) \wedge (\neg P_0(e)) \\
			       & (!e, A, \delta) & \longmapsto & \omega(e, A, \delta) \\
		\end{array}
	\end{equation}
	The \emph{seq} case is the most interesting: the new value for $\delta$ on the right branch is $\delta \wedge P_0(e_1)$. We check for pattern well-formedness in the right branch only if we are supposed to do so ($\delta$ is true) and if the left branch can succeed without consuming any input ($P_0(e_1)$).
		\begin{definition}[Well-formed grammars]
			We can define a well-formed set of grammars $P_{exp}$ as satisfying:
			\begin{equation}\label{compwf}
			\forall A \in V_N, \; \omega(P_{exp}(A), A, \top)
			\end{equation}		
			Such grammars ensure termination of the corresponding parser. 
		\end{definition}
			
			In the PVS implementation, we define the \(\omega\) function (named \texttt{g\_wf}), as well as a \emph{well-formed set of nonterminals} type.


	\section{Abstract Syntax Trees and Parsing}\label{sec:ast}
        In contrast with prior work~\cite{DBLP:conf/popl/Ford04,trx}, we choose not to define parsing as a relationship between an expression, a string and a result, but as a step of a parsing function that corresponds to the actual computation and that verifies a certain number of properties. To do so, we choose to define an output type that represents a computational path, that we call an \emph{abstract syntax tree} as a parse tree that represent the full trace of the parse
        covering both successful and failed branches.  
	This representation
        of the computational path as an explicit proof of correctness
        (soundness and completeness) for our reference parser, makes
        it \emph{easier} to observe and explain both success and
        failure. Unlike the TRX verified PEG parser~\cite{trx}, we
        choose to use this structure and to avoid defining parsing
        through rules.  We believe that this approach is richer, as it provides explicit information on the computational path, and as it is easy to show that the rules of parsing are verified at each node of an Ast tree. An example of a lemma capturing a parsing rule that is verified by the parser is shown in Figure~\ref{pvs:eq_axioms}.
	
	\subsection{Pre-Ast}
	The implementation of those abstract syntax trees will need to verify a certain number of properties in order to represent an actual computational path. But as we can define properties only on existing objects, we start by creating a \emph{pre-Ast} datatype, on which we will build the full \emph{Ast} type. As we can expect, a \emph{pre-Ast} depends on the type of terminals $V_T$, the type of non terminals $V_N$, but also the upper bound\footnote{This can easily be replaced by a notion of \emph{end of stream character} if the input bound is unknown.} of the input $b\in \mathbb{N}$. We have nine cases, that corresponds to nine constructors. Each constructor has its own arguments, but always requires $s,e \in \llbracket 0,b\rrbracket$: the \emph{start} and \emph{end} of the string that was \emph{consumed} by the subtree.
		\begin{enumerate}
			\item $skip(s,e,G)$ with $G \in \Delta$. This corresponds to all the cases where part of the grammar is skipped. For example, in the case of a prioritized choice, the second branch can be \emph{skipped} if the first branch is already a success. For the moment, no condition is added on $s$ and $e$, but later on we will obviously ask for $e = s$. We give the set of skips: $\mathcal{S} = \{T \in \mathcal{A}\; |\; \exists (s,e,G) \in \mathbb{N}\times \mathbb{N} \times\Delta ,\; T = skip(s,e,G)\}$.
			\item $\epsilon(s,e)$ for the corresponding grammar node 
			\item $any(s, e,x)$ for the corresponding grammar node, the the consumed character stored as $x$.
			\item $terminal(s, e, a, x)$ for the corresponding grammar node, the expected character being $a$ and the consumed one being $x$. In the success case, $a=x$ but in the failure case, we want to store exactly why it failed, so we store $x\neq a$.
			\item $nonTerminal(s,e, A, T)$ for the corresponding grammar node, with $T \not\in \mathcal{S}$ being the tree for parsing the non terminal $A$.
			\item $seq(s,e,T_1,T_2)$ with $T_1 \not\in \mathcal{S}$. $T_1$ and $T_2$ are supposed to correspond to the parsing of the sub-expressions $e_1$ and $e_2$.
			\item $prior(s,e,T_1,T_2)$ with $T_1 \not\in \mathcal{S}$. Same as $seq$.
			\item $star(e,s, T_0,T_s)$ with $T_0 \not\in \mathcal{S}$. To keep the star structure, we ask for $T_s$ to be either of type $star$ or $skip$. We define $\mathfrak{S}$ as the set of star-like pre-Ast.
			\item $notP(e, s, T)$ with $T \not\in \mathcal{S}$.
		\end{enumerate}
		Here is the summary:
	\begin{definition}[Pre-Ast]
	We give the following inductive definition: 
		\[
				\begin{array}{rcl}
   \mathcal{A}[V_T, V_N, b]&::= & skip(s,e,G),\ (G\in\Delta) \\
   					&|   &\epsilon(s, e)            \\
					&|	& any(s,e,x),\       	   (x\in V_T)         \\
					&|	& terminal(s,e,a,x),\   (a,x\in V_T)   \\
					&|	& nonTerminal(s,e,A,T),\ (A \in V_N)  \\
					&|	& seq(s, e, T_1, T_2),\ (T_1 \not\in \mathcal{S})  \\
					&|	& prior(s,e,T_1, T_2),\ (T_1 \not\in \mathcal{S})\\
					&|	& star(s, e, T_0, T_s),\ \left(
                                             \begin{array}{l}
                                              T_0 \not\in \mathcal{S},\\
                                               T_s \in \mathcal{S}\cup\mathfrak{S}
                                             \end{array}\right)

   \\
					&|	& notP(s, e, T),\        (T \not\in \mathcal{S})  \\
				\end{array}
		\]
		where, $s,e \in \llbracket 0, b \rrbracket$
	\end{definition}
	As we define the pre-Ast, we see that we want to add extra-conditions on the arguments of the constructors. But some of those conditions rely on a notion of \emph{failure/success}, that we will formalize now.
	\subsection{Failure and Success of Pre-Ast}
	Computing the failure or success requires a depth-first traversal of the tree. We consider three possible outcomes: $\{\bot,\top,u\}$ with $u$ standing for \emph{undefined} and we call this the \emph{type} of a pre-Ast. Here is a description of the function $\eta$ that computes the type:
	\begin{enumerate}
		\item $skip$ is always undefined: the failure/success should not depend on a skip.
		\item $\epsilon$ is a success if $e=s$ (meaning nothing was consumed). Otherwise it is undefined.
		\item $any$ is a success if $e = s +1$, a failure if $e=s$ (meaning the end of string was reached and no character was consumed) or undefined otherwise.
		\item $terminal$ has the same conditions as $any$ and adds the $a=x$ condition.
		\item $nonTerminal$ is of the same type as its subtree
		\item $seq$ is a success if both subtrees are successes, and a failure if $T_1$ fails or if $T_1$ succeeds and $T_2$ fails. All other cases are undefined.
		\item $prior$ is a success if $T_1$ is, or if $T_1$ fails and $T_2$ succeeds. If $T_1$ and $T_2$ fails, it is a failure, and undefined otherwise.
		\item $star$ is always a success, as soon as $T_0$ is not undefined. If $T_0$ is a success, $T_s$ must be a success too (and thus is not a skip)\footnote{This corresponds to the fact that until the search for a pattern $e$ fails, we keep on searching (the star is greedy).}.
		\item $notP$ is the opposite of the subtree type,
                  where the opposite of undefined is also undefined.  
	\end{enumerate}
	\begin{definition}[Meaningful tree]
		A \emph{meaningful} tree is a pre-Ast tree that is either of type $\bot$ or $\top$. If we write $\mathcal{M}$ the set of meaningful trees, we have:
		\begin{equation}
		\mathcal{M} = \eta ^{-1}(\{\bot,\top\})
		\end{equation}
	\end{definition}
	

	\subsection{Well-Formed Tree}
	Once we have a standalone notion of failure and success, we can add the other conditions to make sure the tree is well-formed. Well-formed trees are basically the ones corresponding to a real computational path. Here is a summary of those conditions:
	\begin{enumerate}
		\item $\epsilon$, $any$, $terminal$ are well-formed if they are meaningful.
		\item $nonterminal$ is well-formed if its subtree is and their bounds are equals: $e = e_T$ and $s = s_T$
		\item $seq$: We require $T_1$ to be well-formed, and the bounds of $T_1$ and $T_2$ to be a partition of the bounds of $T$: $s = s_1$, $e_1 = s_2$ and $e_2 = e$. If $T_1$ is a failure, the second part of the grammar is not visited, so $T_2$ must be a skip ($T_2 \in \mathcal{S})$ and it should not consume anything ($s_2 = e_2$).
		\item $prior$: We require $T_1$ to be well-formed,
                  and both $T_1$ and $T_2$ should start at $s$ ($s=s_1$ and $s=s_2$). If $T_1$ is a success, then $T_2$ must be a non-consuming skip, and the end must be the one of $T_1$ ($e = e_1$). If $T_1$ is a failure, the end bound must be the one of $T_2$, that is not allowed to be a skip ($e=e_2$ and $T\not\in \mathcal{S}$)
		\item $star$: We require $T_0$ to be well-formed, and the bounds should be a partition (like with $seq$). If $T_0$ is a success, $T_s$ must be a star, and if $T_0$ is a failure, $T_s$ must be a non-consuming skip.
		\item $notP$ is well-formed if its subtree is. It should not consume anything, so the bounds should be equal: $s = e$.	
		\item $skip$ is always well-formed.
	\end{enumerate}
	We write the set of well-formed trees $\mathcal{W}$. We have the following result.
	\begin{theorem}[Well-formed trees are meaningful]
		A well-formed tree is either a tree of success or of failure:
		\begin{equation}
		\mathcal{W} \subset \mathcal{M}
		\end{equation}
	\end{theorem}
	In PVS, we define the \texttt{astType?} and \texttt{astWellformed} functions over the \texttt{pre\_ast} type. We prove the theorem by induction.
	\subsection{The Definition of Parsing}
		Usually, as we mentioned and as in \cite{DBLP:conf/popl/Ford04} and
                \cite{trx}, the parsing is defined as a relationship
                between the inputs and the outputs that verifies a
                certain number of derivation rules corresponding to all the parsing cases. Lemmas that link the properties of the grammar to the parsing relationship\footnote{Those lemmas are the following : (1) If the parsing fails, the grammar is of type $P_\bot$ (2) If the parsing succeed without consuming any input, the grammar is of type $P_0$ (3) If the parsing succeed consuming at least one token, the grammar is of type $P_{>0}$.} are then proved to ensure that any algorithm following the parsing rules would terminate. We chose not to follow this approach, but to include all the conditions and the lemmas in the typing system of PVS. This has several advantages:
		\begin{itemize}
			\item The typing system is verified by PVS, but once it is proven, it has no impact on the real computation.\footnote{It is also possible to put less information into the type system and prove instead lemmas, but such proofs would need to be done by induction, covering all the possible cases each time. When using the type system, the corresponding induction cases are actually split into the \emph{type-check} conditions.}
			\item We can rely on fewer axioms, as there is no need for a parsing relation-ship to be defined. All the axioms of the parsing relation-ship are proven properties of the parser, and such proofs are easy to do.
		\end{itemize}
		To prove that \emph{well-formed trees} actually correspond to parsing of a given input given a grammar, we introduce two notions. We say that a well-formed tree $T$ is \emph{true to a grammar} $G$\footnote{Here, we are not considering the full set of the nonterminals, but only a grammar node. If this node uses nonterminals, the check of the tree recursively goes into those nonterminals. This notion thus depends on a given set of nonterminals $P_{exp}$} if we can rebuild $G$ from $T$. We say that a well-formed tree $T$ is \emph{true to an input} $I$ if the characters stored in the tree at some starting indices corresponds to the input. This notion only applies to the part of the input \emph{covered } by the tree - namely, between the start and the end indices of the tree\footnote{The full code of the functions \texttt{trueToGrammar} and \texttt{trueToInput} is not given here, at it is only very simple recursive checks. Basically, we do a case analysis on the constructors of the \texttt{pre\_ast} datatype. The code can be seen on the github page.}.  This yields three results that can be proved by induction (see Figure~\ref{pvs:unique} for the PVS code):
		\begin{theorem}[Uniqueness for well-formed trees]
		We have:
		\begin{itemize}
			\item If a grammar $G_1$ and a grammar $G_2$ are both \emph{true} to a given well-formed tree $T$, then $G_1 = G_2$.
			\item If an input $I_1$ and an input $I_2$ are both \emph{true} to a given well-formed tree $T$, and if the starting and ending indexes of the tree are $(s,e)$, then $\forall i \in \llbracket s, e \rrbracket, I_1(i) = I_2(i)$
			\item If two well-formed trees $T_1$ and $T_2$ are both \emph{true} to the same grammar $G$ and to the same input $I$, and start at the same point ($s_{T_1} = s_{T_2}$), then $T_1 = T_2$ 
		\end{itemize}
		\end{theorem}

	\begin{center}
		\begin{figure}[!ht]
\begin{lstlisting}[language=PVS]
% Grammar uniqueness:
unique_grammar: LEMMA ∀(T, G1, G2, P_exp): 
  ( trueToGrammar(T, G1, P_exp) 
  ∧ trueToGrammar(T, G2, P_exp)) 
    ⇒ (G1 = G2)

% Input uniqueness:
unique_input: LEMMA ∀((T: ast), inp1, inp2): 
  ( trueToInput(T, inp1) 
  ∧ trueToInput(T, inp2)) 
    ⇒ (∀ (i:below(e(T))): (i >= s(T)) ⇒ inp1(i) = inp2(i))

% Main uniqueness theorem: 
unique_tree: LEMMA ∀((T1, T2 : ast), inp, G, P_exp): 
  ( trueToInput(T1, inp) 
  ∧ trueToInput(T2, inp) 
  ∧ trueToGrammar(T1, G, P_exp) 
  ∧ trueToGrammar(T2, G, P_exp) 
  ∧ s(T1) = s(T2)) 
    ⇒ T1 = T2
\end{lstlisting}
			\caption{PVS implementation of the uniqueness results.}\label{pvs:unique}
		\end{figure}
	\end{center}
		This result is needed to prove that a parser is \emph{complete}.  Since the parser produces well-formed trees that are \emph{true} to the grammar and input given as arguments,  we get that the result tree is the only one possible. If we add the fact that well-formed trees are always either trees of success or of failure, we get that parsing expression grammars always either succeed or fail, and the resulting tree is unique given the grammar and input.
			In the PVS implementation, we define the \texttt{trueToGrammar} and \texttt{trueToInput} functions and prove the uniqueness results.
\section{A Reference Parser Interpreter}\label{sec:refparser}
	\paragraph{}
	Now that we have a well-defined notion of computational path, along with well-formed grammars, we can define a parser generator function that is surprisingly simple.
	\subsection{Parser Interpreter}
		The $peg\_ parser$ function is thought to be an interpreter for parsers of any PEG.  It takes as input arguments :
		\begin{itemize}
			\item $P_{exp}$ the interpretation of nonterminals. It must be well-formed.
			\item $A\in V_N$ the current nonterminal.
			\item $G\sqsubseteq P_{exp}(A)$ the current grammar node.
			\item $inp$ the input string, represented as an array of characters
			\item $b$ the bound of the parsing, less or equal to the length of the input
			\item $s$ the starting index for the current node
			\item $s_T$ the starting index when the parsing of the current nonterminal started. We ask for $s_T \geq s$ and if $s=s_T$ (which means nothing was consumed since the start of the current nonterminal), we must have $G$ \emph{pattern well-formed}. Indeed, as we saw, we allow subexpressions of a given node $P_{exp}(A)$ to only be {structurally well-formed} only after consumption of a character.
		\end{itemize}
		
		The output type captures a lot of the complexity of the parsing steps, ensuring mostly that trees consumed or failed coherently with the grammar. The output type is the subset of well-formed trees $T$ that verifies the following conditions. (the implementation is given in figure \ref{pvs:outputtype}). The three last conditions ensure that the actual result of the parsing corresponds to a property of the grammar.
		\begin{itemize}
			\item $s(T) = s$
			\item if G is a $star$, then $T$ must be a star tree.
			\item $T\not\in\mathcal{S}$ : T is not a skip.
			\item $T$ is true to the grammar $G$
			\item $T$ is true to the input $I$
			\item $(\eta(T) = \top \wedge e(T) = s(T)) \Rightarrow P_0(G)$ 
			\item $(\eta(T) = \top \wedge e(T) > s(T)) \Rightarrow P_{>0}(G)$ 
			\item $(\eta(T) = \bot) \Rightarrow P_\bot(G)$ 
		\end{itemize}
		
		\begin{figure}[h!]
\begin{lstlisting}[language=PVS]
% Output type : bounds the tree and the grammar
output(
    P_exp: WF_nT, 
    A: below(V_N_b),
    G: {e : Δ | subterm(e, P_exp(A))},
    inp: input,
    s: inp_bound,
    s_T: {k : upto(s) | (k=s) ⇒ g_wf(G, A, P_0c?(P_exp), strong)}) : 
  TYPE = {T : ast | 
    % T starts at the intended position
    (s(T) = s) ∧ 
    % T is true to the grammar and the input
    (trueToGrammar(T, G, P_exp)) ∧
    (trueToInput(T, inp)) ∧ 
    % If the parsing succeeds without consuming something, G is of type P_0
    (((astType?(T) = success) ∧ (e(T) = s)) ⇒ P_0c?(P_exp)(G)) ∧ 
    % If the parsing succeeds consuming something, G is of type P_s
    (((astType?(T) = success) ∧ (e(T) > s)) ⇒ P_sc?(P_exp)(G)) ∧ 
    % If the parsing fails, G is of type P_f
      ((astType?(T) = failure) ⇒ P_fc?(P_exp)(G))}

\end{lstlisting}
			\caption{Output type definition. The \texttt{output} function produces a type based on its arguments. The type is expressed by comprehension over the \texttt{ast} type of well-formed grammars.}
			\label{pvs:outputtype}
		\end{figure}
		
		Those properties ensure that at every step of the computation, the properties needed for termination are preserved. Thanks to all the previous work on grammars and Ast, it is surprisingly simple (see Figure  \ref{pvs:parser}). The types contains most of the valuable information.
	
	\begin{center}
		\begin{figure*}[h!]
\begin{lstlisting}[language=PVS]
parsing(
  P_exp : WF_nT, % Set of nonTerminals
  A: below(V_N_b), % Current non Terminal
  G: {e : Δ | subterm(e, P_exp(A))}, % Grammar node
  inp: input, % Input array
  s: inp_bound, % Starting index 
  % Starting index at the beginning of the parsing of the current non terminal :
  s_T: {k : upto(s) | (k=s) ⇒ g_wf(G, A, P_0c?(P_exp), strong)}) :
RECURSIVE output(P_exp, A, G, inp, s, s_T) =
  CASES G OF
      ε: ε(s,s),
      any: any(s, min(s+1, bound), inp(s)),
      terminal(a): terminal(s, min(s+1, bound), a, inp(s)),
      nonTerminal(B): 
          let T_B = parsing(P_exp, B, P_exp(B), inp, s, s) in 
            nonTerminal(s, e(T_B), B, T_B),
      seq(e1, e2): 
          let T1= parsing(P_exp, A, e1, inp, s, s_T) in
            if (astType?(T1) = failure) then seq(s, e(T1), T1, skip(e(T1), e(T1), e2))
            else 
              let T2 = parsing(P_exp, A, e2, inp, e(T1), s_T) in
                seq(s, e(T2), T1, T2) 
            endif,
      prior(e1, e2):
          let T1 = parsing(P_exp, A, e1, inp, s, s_T) in
            if (astType?(T1) = success) then prior(s, e(T1), T1, skip(s, s, e2))
            else 
              let T2 = parsing(P_exp, A, e2, inp, s, s_T) in
                prior(s, e(T2), T1, T2) 
            endif,
      star(e): 
        let T0 = parsing(P_exp, A, e, inp, s, s_T) in
          if (astType?(T0) = failure) then star(s, s, T0, skip(e(T0), e(T0), star(e)))
          else 
            let Ts = parsing(P_exp, A, star(e), inp, e(T0), s_T) IN
              star(s, e(Ts), T0, Ts) 
          endif,
      notP(e): let T = parsing(P_exp, A, e, inp, s, s_T) IN notP(s, s, T)
  ENDCASES
MEASURE lex4(bound - s_T, bound - s,A, pegMeasure(G))
\end{lstlisting}
			\caption{Code for the parser interpreter. The \texttt{CASES...OF} syntax is used to do a case analysis on the possible constructors for the PEG datatype. As for every recursive function in PVS, we provide the measure ensuring termination (\texttt{lex4} stands for lexical ordering of 4 elements).}
			\label{pvs:parser}
		\end{figure*}
	\end{center}	
	
		\subsection{Termination}
		The termination is proved by using a strictly decreasing lexicographic order on the 4-tuple : \[(b - s_T, b - s, A, |G|)\]
		At each step, we either :
		\begin{enumerate}
			\item Go down the grammar, thus $|G|$ decreases
			\item Use a strictly lower nonterminal, thus $A$ decreases
			\item In the case of a $star$ in the grammar, both the grammar node, the current nonterminal and $s_T$ stay the same, but $s$ increases (because if $e*$ is well-formed, we have $\neg P_0(e)$).
			\item If we get to a nonTerminal node, and that nonTerminal is greater than the current one, the recursive call is made with $s \leftarrow s_T$. Here $s$ must be strictly greater than $s_T$, ensuring that at least one token was consumed before using greater nonTerminal).
		\end{enumerate}
		
		\subsection{Satisfaction of the Parsing Rules}
		As we mentioned, this reference parser satisfies the rules defined by \cite{DBLP:conf/popl/Ford04} by design. This was ensured by structure of well-formed ast-trees, but we can prove it \emph{a posteriori}. It is almost always trivial, as expanding the definition of the parsing functions is enough to show the property. An example is shown in Figure \ref{pvs:eq_axioms}. In the actual implementation, the full set of rules is proven.
	
		\begin{figure}[h!]
\begin{lstlisting}[language=PVS]
parsing_correctness_prior: 
LEMMA FORALL (P_exp : WF_nT, 
  (A : V_N),
  (e1,e2 : {G : Δ | subterm(G, P_exp(A))}),
  (inp : input), 
  (s : upto(bound)), 
  (s_T: {k : upto(s) | (k=s) ⇒ g_wf(prior(e1,e2), A, P_0c?(P_exp), strong)})) :
subterm(prior(e1, e2), P_exp(A)) ⇒
  LET T1 = parsing(P_exp, A, e1,           inp, s, s_T) IN 
  LET T2 = parsing(P_exp, A, e2,           inp, s, s_T) IN 
  LET T  = parsing(P_exp, A, prior(e1,e2), inp, s, s_T) IN
      ((astType?(T1) = ⊤) ⇒ (astType?(T) = ⊤ ∧ e(T) = e(T1))) ∧
      ((astType?(T1) = ⊥) ⇒ (astType?(T) = astType?(T2) ∧ e(T) = e(T2)))

parsing_correctness_notP:
LEMMA FORALL (P_exp : WF_nT, 
  (A : V_N),
  (e : {G : Δ | subterm(G, P_exp(A))}),
  (inp : input),
  (s : upto(bound)),
  (s_T: {k : upto(s) | (k=s) ⇒ g_wf(notP(e), A, P_0c?(P_exp), strong)})) :
subterm(notP(e), P_exp(A)) ⇒
  LET T_n = parsing(P_exp, A, e,       inp, s, s_T) IN 
  LET T   = parsing(P_exp, A, notP(e), inp, s, s_T) IN
    (astType?(T_n) = ⊥ ⇒ (astType?(T) = ⊤ ∧ e(T) = s)) ∧
    (astType?(T_n) = ⊤ ⇒ (astType?(T) = ⊥))
\end{lstlisting}
			\caption{Example of lemmas that shows that the parser verifies the axioms of parsing. In the PVS implementation, the full set of axioms is proven.}
			\label{pvs:eq_axioms}
		\end{figure}
		
		\subsection{Completeness and Soundness}
		As we can see in the output type, the produced trees are true to the input and to the grammar. Thanks to the uniqueness results, this ensures that the only proof of parse or proof of failure is that exists for such a pair of input and grammar is the one found by the parser.

\section{A Packrat Parser Generator}\label{sec:packrat}
The natural extension of a PEG parser is to build a packrat parser (see~\cite{DBLP:conf/icfp/Ford02}) upon the reference parser. As we already have a reference parser, it is easy to build an efficient packrat parser that only relies on the reference one (through the typing system). The output tree can also be compacted on the fly, removing useless failing branches, while ensuring that the output is \emph{equivalent} to the one of the reference parser. This allows us to use the reference parser as a \emph{ghost} reference while never actually calling it.
This whole section is directly inspired by the work of Ford~\cite{DBLP:conf/icfp/Ford02}.
	
\subsection{Complexity}

The key to transform the worst case exponential complexity of the parsing into a linear one comes from the following observations:
\begin{itemize}
	\item Searching for a pattern $A$ at a starting position $s$ is independent from previously done parsing, it only relies on $A$ and $s$.
	\item There are at most $n$ patterns we can search at most $b$ starting positions		
\end{itemize}
We deduce that if we store the result of the parsing of nonterminals at given starting positions when we compute them (laziness), the parser will be called at most $b\times n$ times. As a recursive call is made in constant time\footnote{In the reference parser implementation, the use of \lstin{astType} function makes it non-constant, but this can be easily changed by modifying the parser so it also outputs the type of the return tree. This is done in the compacted semantic trees of section 6, as failure trees are reduced to failure nodes, hence making typechecking of trees trivial.}, we end up with a complexity bounded by $O(b\times n)$.

\subsection{Memoization}

We use a PVS structure to store results as they are computed, as shown in Figure~\ref{pvs:result}. The type of those objects is based on the reference parser, ensuring that we only store results that we could obtain by calling the parser with the same parameters. We then modify the parsing function to return both the ast and the record of computed results. When we parse a nonterminal, the result is either already known (and directly returned), or we compute it and update the record of results. This modification is shown in Figure~\ref{pvs:packratnonterm}.

\begin{figure}[!h]
\begin{lstlisting}[language=PVS]
% Datatype used for memoization of partial parses
saved_result: DATATYPE
BEGIN
  unknown : unknown?
  known(T : ast) : known?
END saved_result

% Results type : ensures that every partial parse correspond to the reference parser
results(
    P_exp: WF_nT,
    inp: input) : 
  TYPE =
    {r : [V_N -> [inp_bound -> saved_result]] | 
    FORALL (A : V_N, s : upto(bound)) : 
      known?(r(A)(s)) ⇒ T(r(A)(s)) = parsing(P_exp, A, P_exp(A), inp, s, s)}
\end{lstlisting}
\caption{The result structure to store intermediate results as they are computed. A datatype with two constructors is used. The \texttt{results} function produces a type by comprehension over the set of functions which take a nonterminal and a starting index and returns a stored result. Such a stored result should be equal to the reference parser called on the same arguments.}
\label{pvs:result}
\end{figure}

\begin{figure}[!h]
\begin{lstlisting}[language=PVS]
CASES G OF 
  ...
  nonTerminal(B): (
      CASES res(B)(s) OF
        known(T_B): (nonTerminal(s, e(T_B), B, T_B), res),
        unknown: 
          let (T_B, resB) = packrat_parser(P_exp, B, P_exp(B), inp, s, s, res) in
            (nonTerminal(s, e(T_B), B, T_B), resB WITH [B := resB(B) WITH [s := known(T_B)]])
      ENDCASES),
  ...
ENDCASES
\end{lstlisting}
\caption{Parsing a nonterminal, checking if the result is already known. If the result is unknown, we compute it and extend the function using the \texttt{WITH [.. := ..]} syntax.}
\label{pvs:packratnonterm}
\end{figure}

\subsection{Change of the Output Type}

As we mentioned, the reference parser can be used as a \emph{reference}, stating that the results of the packrat parser are the same as if the reference parser was called with the same arguments. This is illustrated in Figure~\ref{pvs:packratoutput}. Again, those conditions are typing conditions, which means that once they are proved through the typechecking condition system, they do not impact the actual computation.

\begin{figure}[!h]
\begin{lstlisting}[language=PVS]
packrat_parser(
  P_exp: WF_nT, A: below(V_N_b),
  G: {e : Δ | subterm(e, P_exp(A))},
  inp: input, s: upto(bound),
  s_T: {k: upto(s) | (k=s) ⇒ g_wf(G, A, P_0c?(P_exp), strong)},
  res: results(P_exp, inp)) :
  RECURSIVE 
    [
      {T : pre_ast | T = parsing(P_exp, A, G, inp, s, s_T)}, 
      results(P_exp, inp)
    ]
\end{lstlisting}
\caption{The packrat parser returns a tree that is the same as the reference parser, and a function that satisfies the \texttt{result} type.}
\label{pvs:packratoutput}
\end{figure}
\section{Semantic Compacted Tree}\label{sec:semantic}

In this section we  introduce a modification of the parser that allows the user to specify the data-structure of the parsed result according to their specific needs. Indeed, the \lstin{AST} we defined are interesting as effective proofs of parse\footnote{Especially since they provide a complete description of the computational path of the parser.} but in practice, users are more interested in data-structures specific to their grammar and not to the underlying PEG operators. For example, an HTML parser should output a DOM tree and not the cumbersome full \lstin{AST} tree of the parser generator. In order to modify the parser to be able to create those semantic trees we  need to introduce modifications to the \lstin{AST} type, specify how the user provides the constructors for their data structure, and how we maintain the equivalence with the reference parser.

\subsection{Semantic/Failure Node}
To allow the user to produce custom data-structures, we need to introduce a new type for those data-structures : $V_S$. In order to simplify the definitions, we do not distinguish between the subtypes the user may be using in the data structure. We only consider that $V_S$ is the superset of all user types. We call elements of $V_S$ semantic values. When designing a grammar, the user creates nonterminals for each pattern in the intended input. Therefore, the meaningful unit that can be transformed into a semantic value is the result of the parsing of a nonterminal. To replace nonterminal \lstin{AST}s by their semantic value, we need to add a new constructor to the \lstin{AST} type to store such values. This is shown in Figure~\ref{pvs:semanticnode}. Following the same idea, we also introduce a new failure node to replace failing trees that are useless to the user.

\begin{figure}[h!]
\begin{lstlisting}[language=PVS]
pre_ast : DATATYPE
	...
	% s : start index
	% e : end index
	% A : non terminal that was interpreted as semantic value
	% S : semantic value stored
	semantic(s, e: below, A: V_N, S: V_S): semantic?
	% s : start index
	% e : end index
	fail(s, e): fail?
END pre_ast
\end{lstlisting}
	\caption{New constructors for the \lstin{AST} type. Those two new constructors are used to compact the tree.}
	\label{pvs:semanticnode}
\end{figure}

\subsection{Semantic Interpretation of an \lstin{AST}}
\begin{figure*}[h!]
\begin{lstlisting}[language=PVS]
% Function that transforms a tree into its semantic interpretation, recursively
s_inp(
    P_inp: semantic_interp,
    T: pre_ast): 
  RECURSIVE 
    {T': semanticTree | 
      ( (skip?(T) ⇔ skip?(T') )
      ∧ (star?(T) ⇔ star?(T') )
      ∧ ((astType?(T) = failure) ⇔ fail?(T') )
      ∧ (s(T') = s(T))
      ∧ (e(T') = e(T))
      ∧ (astWellformed?(T) ⇒ (astWellformed?(T') 
      ∧ (astType?(T') = astType?(T))))} =
  IF (astType?(T) = failure) THEN fail(s(T), e(T)) 
  ELSE
    CASES T OF
        ε(s, e): T,
        any (s, e, x): T,
        terminal (s, e, x, y): T,
        nonTerminal(s, e, A, T): semantic(s, e, A, P_inp(A, s_inp(P_inp, T))),
        semantic (s, e, A, S): T,
        seq(s, e, T1, T2): seq(s, e, s_inp(P_inp, T1), s_inp(P_inp, T2)),
        prior(s, e, T1, T2): prior(s, e, s_inp(P_inp, T1), s_inp(P_inp, T2)),
        star (s, e, T0, Ts): star (s, e, s_inp(P_inp, T0), s_inp(P_inp, Ts)),
        notP(s, e, T): notP (s, e, s_inp(P_inp, T)),
        skip(s, e, G): T
    ENDCASES
  ENDIF
MEASURE astMeasure(T)
\end{lstlisting}
	\caption{Recursive semantic interpretation of a tree.}
	\label{pvs:semantic}
\end{figure*}

\begin{figure}[h!]
	\includegraphics[width=0.3\textwidth]{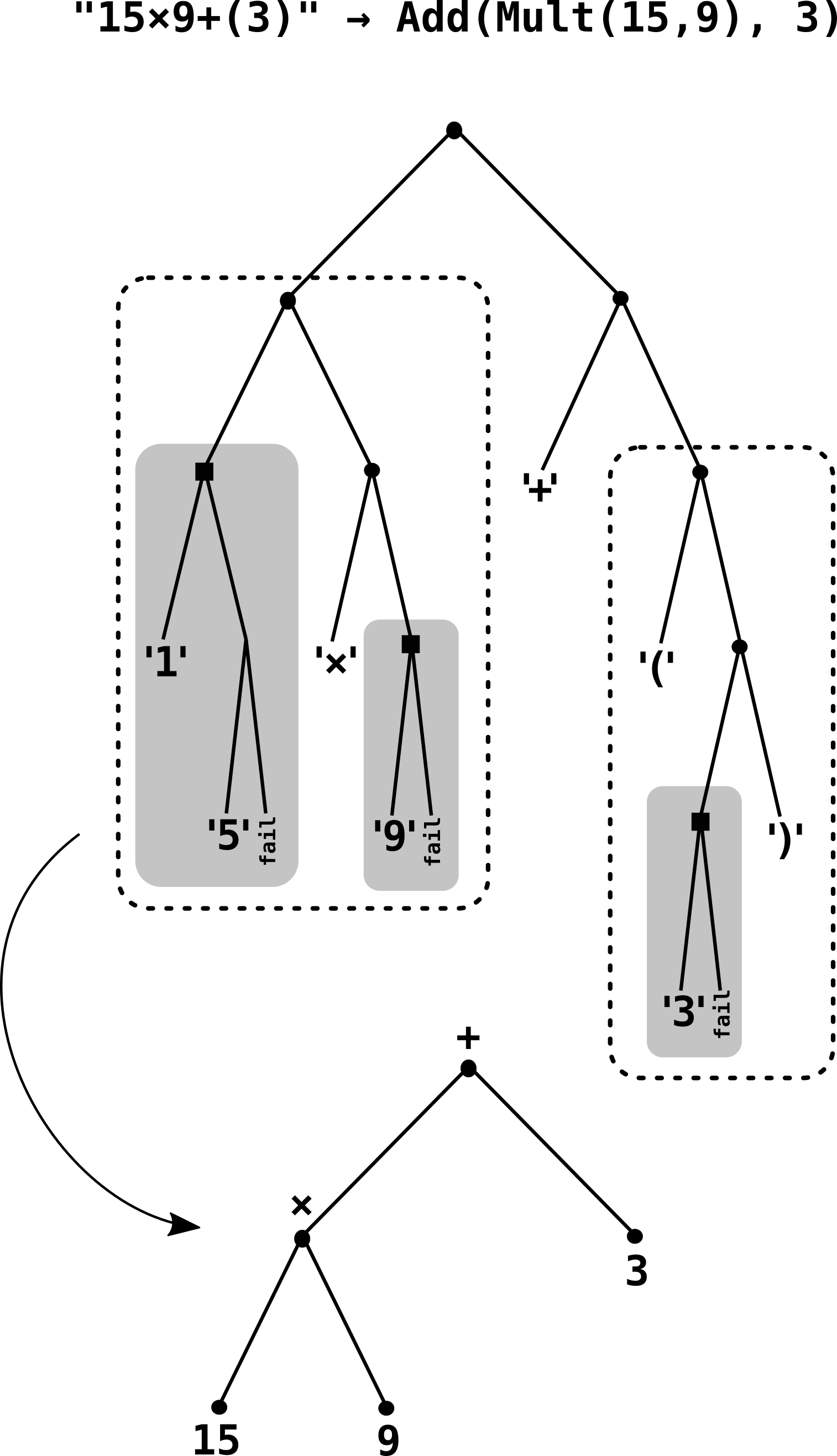}
	\caption{Representation of the semantic interpretation of simple arithmetic expression. Numbers are parsed as lists of digits (grey boxes).}
	\label{png:semantic}
\end{figure}

Now that we can store semantic values (that correspond to nonterminal
subtrees), we can replace those trees by their \emph{computed semantic
  value}. This transformation is called the \emph{semantic interpretation
  of a tree}. It can be done on a fully computed \lstin{AST} produced
by the parser, or on the fly while parsing\footnote{Even though the function can compute semantic interpretations on full trees, the idea is to do it on the fly, and use the function on full tree only to prove equivalence with the reference parser. At runtime, in order to store the smallest possible trees (that can get very big for complex grammars), we compact subtrees as soon as possible by taking their semantic interpretations}. The user provides a function that does the transformation of an nonterminal node ($P_{inp}
: AST \longrightarrow V_S$) and we do a depth-first search in the tree to
recursively replace those nodes by their semantic value. To compact
the tree, the semantic interpretation also replaces failing tree by a
simple \lstin{fail} node. The implementation is given in
Figure~\ref{pvs:semantic}.\footnote{A few conditions are added to the
  output type for the sake of simplification later on in the parser.} A tree that has no nonterminal nodes (but semantic ones) and no failing branches (except for fail nodes) is called a \emph{semantic compacted tree}. A \emph{semantic parser} will simply compute the semantic interpretation of the nonterminal on the fly during the parsing. A representation of the semantic interpretation of a simple arithmetic expression is given in Figure~\ref{png:semantic}.

\subsection{Semantic Equivalence with Reference Parser \lstin{AST}}
The notion of equivalence with the reference parser becomes more subtle here, as compacted semantic trees are not equal to the trees produced by the reference parser. The main idea is the following: we want to ensure that when information is condensed in a compacted tree, we get the same information if we call the reference parser on the same input and then compact the resulting tree. Here is the detailed definition of equivalence:
\begin{enumerate}
	\item For all nodes that are not \lstin{fail} and \lstin{semantic} nodes, the reference and compacted trees should be the same.
	\item For a \lstin{fail(s, e)} node, we need to check that if we call the reference parser on the same starting point, we would get a failing tree ending on the same index \lstin{e}.
	\item For a \lstin{semantic(s, e, A, S)} node, we need to check that the semantic interpretation of the output of the reference parser would be equal to the value stored here.
\end{enumerate}
These conditions on the output of the semantic parser interpreter can be captured directly in the typing system.
The output type of a semantic parser interpreter is the following (using the \texttt{s\_inp} function). 
\begin{lstlisting}[language=PVS]
{T: pre_ast | T = s_inp(P_inp, parsing(P_exp, A, G, inp, b, s, s_T))}
\end{lstlisting}

\paragraph{Example.}
The PVS suite provides an interactive interface to execute the verified code: PVSio~\cite{DBLP:conf/tap/DutleMNB15}\@.
Using this system, we can test the parser on real examples. We
crafted a very simple arithmetic expression parser for test purposes:
the expression is represented as an array of ascii values
($1+2*(3-4/5)$ is represented as the sequence $[49, 43, 50, 42, 40, 51, 45, 52, 47,
53, 41]$ with binary values). The interaction with the system is shown
in Figure~\ref{pvsio:sum}. We can see that the semantic parser does produce a compacted tree that correspond to the reference one, and that modifying the input changes the result accordingly.
\begin{figure}[h!]
\begin{lstlisting}[language=PVS]
% Parsing of the arithmetic expression 1*2+(3-4/5)
% (encoded in ascii : 49 42 50 43 40 51 45 52 47 53 41)

<PVSio> parser_semantic(inp1);
==>
semantic(0, 11, 8, add(mult(v(1), v(2)),
                       sub(v(3), div(v(4), v(5)))))

% If we twist the '+' into a '-' we have :
<PVSio> parser_semantic(inp4 WITH [3 := 45]);
==>
semantic(0, 11, 8, sub(mult(v(1), v(2)),
                       sub(v(3), div(v(4), v(5)))))

% If we introduce an error : 
<PVSio> parser_semantic(inp1 WITH [1 := -3]);
==>
fail(0,1)
\end{lstlisting}
	\caption{Example of using the semantic parser for simple arithmetic expressions}
	\label{pvsio:sum}
\end{figure}

\section{Conclusions}\label{sec:conclusions}

Extracting data from input streams representing
programs, text, documents, images, and video is a complex task.
Parsers for these data formats transform the
input data streams into actionable data while rejecting incorrect inputs.
Parsing is supported by a rich body of theory, but it is also 
central to practice.  
Many software vulnerabilities arise from poorly designed grammars,
ambiguous inputs, and incorrect parsing.  Parsing expression grammars
are a widely used class of expressive grammars that support efficient
and unambiguous packrat parsing through memoization.  We have used PVS to formalize
the metatheory of PEGs and derived a correct parser
interpreters for these grammars supporting memoization and semantic actions.
The proofs have been mechanically verified in PVS
by taking advantage of the expressiveness of the PVS type
system.  A significant part of the metatheory covers
 the analysis of grammars constructed from PEG expressions to check if a
 parse based on the expression could fail, could succeed without consuming input,
 or could succeed only by consuming input.  This analysis is used to establish the
 termination of a reference parser interpreter for PEGs.  We also define a
 parse tree representation that captures the trace of the parser on an input.
 This representation serves as a evidence for the parsing of the input with respect
 to a given grammar.  We establish
 certain uniqueness results that demonstrate that there is exactly one grammar
 with respect to parse tree for a given input, and exactly one parse tree for a
 given input with respect to a given grammar.  These uniqueness results demonstrate
 that the reference parser is complete: it either returns a successful parse tree
 and there is a unique such tree, or a failed parse tree and there is no other
 parse tree that parses the same input with success or failure.  

 In future work, we plan to conduct empirical studies of
 the performance of the parsers, explore the use of efficient data
 structures for parsing, develop a systematic methodology for the
 derivation of correct-by-construction parser interpreters and
 generators for other grammar formats, and experiment with the
 integration of the generated parsers into security-critical
 applications.  Our work is a preliminary step toward powerful correct-by-construction
parser interpreters and generators 
 for expressive grammars and data format
 descriptions.


\balance
\bibliography{biblio.bib}


\end{document}